\documentclass[12pt]{iopart}
\usepackage{xcolor}
\usepackage{hyperref}
\usepackage{braket}
\usepackage{amsfonts}
\expandafter\let\csname equation*\endcsname\relax
\expandafter\let\csname endequation*\endcsname\relax
\usepackage{amsmath}
\usepackage{graphicx}
\usepackage{comment}
\usepackage{mathtools}
\usepackage{cite}

\begin{document}

\title{Work fluctuations for a confined Brownian particle: the role of initial conditions}

\author{Giovanni Battista Carollo$^{1*}$, Massimiliano Semeraro$^{1}$, Giuseppe Gonnella$^1$ and Marco Zamparo$^1$}

\address{$^1$Dipartimento di Fisica, Universit\`a degli Studi di Bari and INFN, Sezione di Bari, via Amendola 173, 70126 Bari, Italy}
\ead{* \href{mailto:giovanni.carollo@uniba.it}{giovanni.carollo@uniba.it}}
\vspace{10pt}
\begin{indented}
\item[]July 2023
\end{indented}

\begin{abstract} 
We study the large fluctuations of the work injected
  by the random force into a Brownian particle under the action of a
  confining harmonic potential. In particular, we compute analytically
  the rate function for generic uncorrelated initial conditions,
  showing that, depending on the initial spread, it can exhibit no,
  one, or two singularities associated to the onset of linear tails. A
  dependence on the potential strength is observed for large initial
  spreads (entailing two singularities), which is lost for stationary
  initial conditions (giving one singularity) and concentrated initial
  values (no singularity).  We discuss the mechanism responsible for
  the singularities of the rate function, identifying it as a big jump
  in the initial values.  Analytical results are corroborated by
  numerical simulations.

\end{abstract}

\section{Introduction}
\label{sec:intro}
Rare events represent an active research field in physics,
mathematics and natural sciences \cite{albeverio2006,
  gumbel2012}. One way to describe them is suggested by large
deviation theory \cite{dembo1984, denhollander2000}, which provides a
quantitative evaluation of probabilities of events beyond the regime
of normal fluctuations. This theory also offers the possibility to
extend the usual equilibrium free-energy approach of statistical
physics to non-equilibrium and dynamical contexts
\cite{touchette2009}. In fact, given an extensive physical observable
$W_{\tau}$ computed by cumulating a large number $\tau$ of microscopic
events, if a large deviation principle holds, then the asymptotics of
the probability distribution $P(W_\tau/\tau=w)$ can be characterised
by the rate function
$I(w)=-\lim_{\tau\uparrow\infty}\frac{1}{\tau}\log P(W_\tau/\tau=w)$,
playing indeed a role similar to a free-energy.

A singularity in the rate function can separate the regime of typical
fluctuations around the mean from the regime of far rare events. In
dynamical contexts, singularities of this kind are associated with
dynamical phase transitions \cite{jack2020, garrahan2009, jack2010,
  lefevere2011, speck2012, corberi2013, szavits2014, zannetti2014,
  touchette2016,corberi2019, gradenigo2019, zamparo2019,
  zamparo2022}. In particular, if $W_{\tau}$ is an observable measured
along the trajectories of a given system, a singularity in the graph
of the rate function $I(w)$ can mark a phase separation in trajectory
space. A relevant question is which are the trajectories contributing
to the different regimes of $I(w)$. A possible answer could come from
the so-called single big-jump principle
\cite{bigjumpbook,vezzani2019,burioni2020,my_bigjump}, which explains
rare events not in terms of an accumulation of many small microscopic
events but solely as an effect of the biggest one. It would be
interesting to analyse this principle in the context of dynamical
phase transitions.

The dynamics of Brownian particles offers the possibility to deepen
our understanding of the above subjects.  Ref.\ \cite{farago2002}
studied the work done by the thermal bath force on a single
underdamped Brownian particle. In particular, the rate function of the
work for a free particle was computed and, in the stationary regime, a
singularity and an associated linear tail were found at negative
values of the work. The consequences on the non validity of the
Fluctuation Relation were discussed. Then, the effects of a confining
potential were considered.  A confining potential at the level of a
single particle, in addition to having an intrinsic interest as
experimentally realisable by means of an optical trap, can mimic the
trapping created by other particles at finite densities, and has been
studied with also this purpose \cite{szamel2014, nandi2017,
  woillez2020}.  Arguments were given in Ref.\ \cite{farago2002} to
conclude that the presence of a harmonic potential would not modify
the rate function for the thermal bath work with respect to the free
case. Singular rate functions for trajectory dependent quantities have
also been found for Brownian particles in a moving potential
\cite{van2003, pal2013} or in contact with several baths
\cite{visco2006}, for Brownian particles under the
  action of an additional Gaussian force \cite{sabhapandit2011,
    arnab2014}, for single harmonically confined active particles
\cite{semeraro2023} and for active Brownian particles at finite
densities \cite{cagnetta2017, nemoto2019, fodor2020, chiarantoni2020,
  keta2021}.

In this paper, we consider again a harmonically confined Brownian
particle and compute analytically the rate function of the work done by the
random force of the thermal bath.  We extend the analysis to the case
of generic uncorrelated initial conditions.  Our results prove the
claims made in Ref.\ \cite{farago2002}, as they show that the rate function does
not depend on the strength of the potential in the stationary regime
and for concentrated initial values. For generic initial conditions,
we demonstrate that the rate function always possesses a singularity at a value
of the work smaller than the mean work, except for the limiting case
of concentrated initial values where such singularity moves to the
boundary. This singularity corresponds to the singularity observed in
Ref.\ \cite{farago2002} for the stationary, free particle.  In
addition, we demonstrate that a second singularity emerges above the
mean work when the spread of the initial conditions is large
enough. Each singularity is the beginning point of an associated
linear tail in the graph of the rate function.  We provide an interpretation of
the singularities in terms of the single big-jump principle, showing
that trajectories in the linear tail regimes are characterised by a
big jump in the initial values.

The paper is organised as follows. In Section \ref{sec:mod_met} we
present the model and introduce the work injected by the random force.
In this section, we also report a brief summary of the analytical
approach developed in Ref.\ \cite{zamparo2023}, which we use to
compute the rate function.  Section \ref{sec:rate_function} is devoted
to the computation of the scaled cumulant generating function (SCGF)
and the rate function, the latter being the Legendre transform of the
former.  In Section \ref{sec:sec-sings} we investigate on the
trajectories phenomenology in order to explain the mechanism that
causes the singularities of the rate function.  Finally, Section
\ref{sec:conclusions} summarises our findings and
  future perspectives.

\section{Model and Methods}
\label{sec:mod_met}

\subsection{Definition of the model}
\label{sec:model}
We consider a single unit-mass Brownian particle in one dimension
under the action of a harmonic potential described by the following
Langevin equation
\begin{equation}
    \ddot x(t)=-\gamma \dot x (t)-kx(t)+\sqrt{2D}~\eta(t),
    \label{eq:ode}
\end{equation}
where $x(t)$ is the particle position, $\gamma$ is the viscous
friction coefficient, $k$ (the potential strength) is the elastic
constant, $\eta(t)$ (the random force) is a Gaussian white noise with
$\braket{\eta(t)}=0$ and $\braket{\eta(t) \eta(s)}=\delta(t-s)$ and
$D$ is the diffusion coefficient. Following Ref.\ \cite{farago2002},
we interpret $-\gamma\dot x(t)$ and $\sqrt{2D}~\eta(t)$ as generic
energy dissipation and injection channels, respectively, so that
throughout calculations we keep $D$ unspecified. The usual case of an
equilibrium thermal bath at temperature $T$ is recovered by fixing
$D=\gamma k_BT$ with $k_B$ the Boltzmann constant, as prescribed by
the Einstein relation.  Introducing the velocity $v(t)$,
Eq.\ (\ref{eq:ode}) is recast into the following system of first-order
differential equations
\begin{equation}
\begin{cases}
\dot x(t)=v(t)\\
\dot v(t)=- \gamma v(t) -kx(t)+\sqrt{2D}~\eta(t)
\end{cases},
\label{eq:systemeom}
\end{equation}
with initial conditions $x(0)$ and $v(0)$, to which we will refer as
the equations of motion.  We consider uncorrelated Gaussian initial
conditions with zero mean and standard deviations $\sigma_x$ for the
position and $\sigma_v$ for the velocity. The case of a stationary
process is obtained with $\sigma^2_x=D/k\gamma$ and
$\sigma^2_v=D/\gamma$ \cite{gardiner2009}.

We will study the probability distribution of the work injected by the
random force into the Brownian particle up to time $\tau$, defined as
\begin{eqnarray}
  \nonumber
W_\tau&\equiv\sqrt{2D}\int _0^\tau \eta(t)\dot x(t)~dt\\
      &=\frac{1}{2}\big[v^2(\tau)-v^2(0)\big]+\frac{k}{2}\big[x^2(\tau)-x^2(0)\big]+\gamma\int_0^\tau v(t)\dot{x}(t)~dt,
\label{eq:work1}
\end{eqnarray}
where in the second row we have used the $\eta(t)$ expression
extracted from Eq.\ (\ref{eq:systemeom}). Our goal is in fact to
compute analytically the rate function $I(w)$ of the work injected per unit
time:
\begin{equation*}
 I(w)\equiv-\lim_{\tau\uparrow\infty}\frac{1}{\tau}\log P\bigg(\frac{W_\tau}{\tau}=w\bigg),
\end{equation*}
where $P(W_\tau/\tau=w)$ is the probability distribution expressed by
the path integral
\begin{equation*}
P\bigg(\frac{W_\tau}{\tau}=w\bigg)=\int \mathcal P_\tau\,\delta(W_\tau-w\tau)\mathcal ~\mathcal Dx\mathcal Dv
\end{equation*}
with path probability 
\begin{equation*}
  \mathcal P_\tau\propto e^{-\frac{1}{2}\big[\frac{x(0)}{\sigma_x}\big]^2-\frac{1}{2}\big[\frac{v(0)}{\sigma_v}\big]^2}
  e^{-\frac{1}{4D}\int_0^\tau[ \dot v(t)+\gamma v (t)+kx(t)]^2dt}\delta(v(t)-\dot{x}(t)),
\end{equation*}
which weighs each trajectory realisation combining the distribution of
the initial values with the Onsager-Machlup weight
\cite{OnsagerMachlup}. To compute the rate function $I(w)$, we follow
the approach developed in Ref.\ \cite{zamparo2023}, which is briefly
summarised in the next subsection. Application of this approach to our
problem is made in Section \ref{sec:rate_function}.

We mention that an alternative approach to deal with
  the work injected by a Gaussian external force into an underdamped
  Brownian particle is suggested in
  Refs.\ \cite{pal2013,sabhapandit2011,arnab2014}. Although both
  methods are ultimately traced back to Gaussian integrals for
  computational purposes, the theory developed in
  Ref.\ \cite{zamparo2023} provides, with mathematical rigour, a large
  deviation principle for a general class of quadratic functionals and
  a comprehensive recipe for evaluating their rate functions.

\subsection{Overview of the analytical approach}
\label{sec:anal_overview}
In order to evaluate the rate function $I(w)$, we resort to the approach recently
proposed in Ref.\ \cite{zamparo2023} for quadratic functionals of
stable Gauss-Markov chains, whose applicability to our problem is
feasible upon a convenient time discretization.  Results for the
original problem are later obtained by performing a continuum limit.
Let $\{X_n\}_{n\geq0}$ be a Markov chain taking values in $\mathbb
R^d$ (in our case $d=2$, corresponding to position and velocity),
and assume that there exist a drift matrix $S$ with spectral radius
$\rho(S)<1$ and an invertible, diagonal diffusion matrix $C$ such that
\begin{equation}
X_{n+1}=S X_n+C G_n,
\label{eq:chain}
\end{equation}
where $\{G_n\}_{n\geq0}$ is a sequence of i.i.d.\ standard Gaussian
random vectors valued in $\mathbb R^d$. Suppose that the initial state $X_0$
is a Gaussian random vector independent of $\{G_n\}_{n\geq0}$ with
zero mean and positive-definite covariance matrix $\Sigma_0$. The
chain (\ref{eq:chain}) is stationary if and only if
$\Sigma_0=\Sigma_s\equiv \sum_{m\geq0}S^m C^2(S^\top)^m$, $\top$
denoting transposition, and the request on the spectral radius ensures
that $\Sigma_s$ actually exists.  Let also
\begin{equation}
W_N=\frac{1}{2}\Braket{X_0, LX_0}+\frac{1}{2}\sum_{n=0}^N\Braket{X_n, UX_n}+\frac{1}{2}\Braket{X_N, RX_N}+\sum_{n=1}^N\Braket{X_n, VX_{n-1}} \\
\label{eq:quadfunct}
\end{equation}
be a quadratic functional of the chain, where
$\langle\cdot,\cdot\rangle$ is the Euclidean scalar product and $L$,
$U$, $R$, and $V$ are $d\times d$ real matrices with $L$, $U$, and $R$
symmetric. With full probability, the typical value of $W_N$ is given
by the following law of large numbers \cite{zamparo2023}
\begin{equation}
\lim_{N\uparrow\infty}\frac{W_N}{N}=\frac{1}{2}\text{tr}\left[\big(U+V^\top S+S^\top V\big)\Sigma_s\right].
\label{eq:mathave}
\end{equation}

The theory developed in Ref.\ \cite{zamparo2023} describes the large
deviations of $W_N$. To expose it, we need to formulate some
mathematical details. For each real numbers $\mu$ and $\theta$ we
introduce the Hermitian matrix
\begin{equation}
F_\mu(\theta)=C^{-2}+  S^\top C^{-2}  S-\mu   U-(C^{-2}   S+\mu   V)e^{-\imath\theta}-(C^{-2}   S+\mu   V)^\top e^{\imath\theta},
\label{eq:mathsymbol}
\end{equation}
where $\imath$ is the imaginary unit, and we term {\it primary domain}
the set $O$ of $\mu$ for which $F_\mu(\theta)$ is positive definite
for all $\theta$. It turns out that $O$ is an interval
$(\tilde\mu_-,\tilde \mu_+)$ with $\tilde\mu_\pm$ extended real
numbers,
i.e., $\tilde\mu_\pm\in\mathbb{R}\cup\{-\infty,+\infty\}$. For $\mu\in
O=(\tilde\mu_-,\tilde \mu_+)$, it is possible to define the integrals
\begin{equation}
\varphi(\mu)\equiv-\frac{1}{4\pi}\int_0^{2\pi}d\theta \log\det F_\mu(\theta)-\log\det C,
\label{eq:math_SCGF}
\end{equation}
\begin{equation}
\Phi_\mu(n)\equiv\frac{1}{2\pi}\int_0^{2\pi}d\theta~e^{-\imath n \theta}F_\mu^{-1}(\theta),\quad n\in\mathbb{Z}.
\label{eq:phiexp}
\end{equation}
The matrices 
\begin{equation}
\begin{aligned}
H_\mu&\equiv I+(C^{-2}S+\mu  V)\Phi_\mu(1), \\
K_\mu&\equiv I+\Phi_\mu(1)(C^{-2}S+\mu  V),
\end{aligned}
\label{eq:mathHK}
\end{equation}
with $I$ the identity matrix, can be shown to be invertible
\cite{zamparo2023} and
\begin{equation}
\begin{aligned}
\mathcal L_\mu&\equiv \Sigma_0^{-1}+  S^\top C^{-2}  S-\mu   L-(  S^\top C^{-2} +\mu   V^\top)  \Phi_\mu(0)   H^{-1}_\mu (C^{-2}  S +\mu   V),\\
\mathcal R_\mu&\equiv C^{-2}-\mu   R-(C^{-2}  S +\mu   V)  K_\mu^{-1}  \Phi_\mu(0)(  S^\top C^{-2} +\mu   V^\top)
\end{aligned}
\label{eq:mathLR}
\end{equation}
can be proved to be Hermitian \cite{zamparo2023}. The interval
$E\equiv(\mu_-,\mu_+)\subseteq O$ in which $\mathcal L_\mu$ and
$\mathcal R_\mu$ are simultaneously positive definite is termed {\it
  effective domain}. We are now in the position to state the following
large deviation result \cite{zamparo2023}.  The SCGF of $W_N/N$ in the
large $N$ limit, in symbols $\lim_{N\uparrow\infty}\frac{1}{N}\ln
\Braket{e^{\mu W_N}}$, turns out to be the function
(\ref{eq:math_SCGF}) with domain $E$ fulfilling
\begin{equation*}
    -\infty\leq\tilde\mu_-\leq\mu_-<0<\mu_+\leq\tilde\mu_+\leq+\infty.
\end{equation*}
Moreover, the quadratic functional $W_N/N$ satisfies a large deviation
principle with rate function $J(w)$ given by the Legendre transform of
the function (\ref{eq:math_SCGF}) in $E$, i.e.
\begin{equation}
    J(w)=\sup_{\mu\in E}\big\{w\mu-\varphi(\mu)\big\}.
\label{eq:discrete_RF}
\end{equation}

Within the effective domain $E$, the SCGF is differentiable, and the
limits $\lim_{\mu\downarrow \mu_-}\varphi'(\mu)\equiv w_-$ and
$\lim_{\mu\uparrow \mu_+}\varphi'(\mu)\equiv w_+$ exist by
convexity. When $\tilde\mu_-<\mu_-$ and/or $\mu_+<\tilde\mu_+$, the
SCGF is non-steep at the boundaries, i.e., $w_\pm$ are finite, and the
general expression for the corresponding rate function is
\begin{equation*}
  J(w)=\begin{cases}
  w\mu_--\varphi(\mu_-) & \mbox{if }\quad w\le w_-\\
j(w)                    & w_-<w<w_+\\
  w\mu_+-\varphi(\mu_+)  & \mbox{if }\quad w\ge w_+
  \end{cases},
\end{equation*}
where $j(w)$ is a convex function connecting continuously at $w_\pm$
along with its first derivative. The rate function is characterised by
second-order singularities at $w_\pm$ and linear tails outside the
interval $(w_-, w_+)$. Thus, whenever the effective domain is smaller
than the primary one, at least one linear stretch in the rate function occurs. We
stress that the matrices $\Sigma_0$, $L$, and $R$ neither affect the
typical value of $W_N$ nor the SCGF function expression and the
primary domain. Instead, they play a crucial role in determining the
effective domain extension, and hence in determining the occurrence of
singularities and linear tails in the graph of the rate function.

\section{Scaled cumulant generating function and rate function}
\label{sec:rate_function}

In this section we apply the above method to our problem.  Subsection
\ref{sec:discretization} introduces a convenient time discretization.
Subsection \ref{sec:scgf} gives the SCGF, whereas the effective domain
is discussed in Subsection \ref{sec:eff_dom}. Finally, the rate function is
investigated in Subsection \ref{sec:sec_rf}. The reader interested
only in the latter can directly go over to the last subsection.

\subsection{Discretization}
\label{sec:discretization}

By time discretization, we recast the equations of motion
(\ref{eq:systemeom}) and the work (\ref{eq:work1}) as a Gauss-Markov
chain and a quadratic functional, respectively. To make a contact with
the framework of Ref.\ \cite{zamparo2023}, time discretization must
provide an invertible, diagonal diffusion matrix $C$. To this aim
we introduce a fictitious noise in the upper equation of
Eq.\ (\ref{eq:systemeom}) with vanishing diffusion coefficient in the
continuum limit. Specifically, we divide the time interval of duration
$\tau$ in Eq.\ (\ref{eq:work1}) into the sum of $N$ time steps of size
$\epsilon$, so that $\tau=N\epsilon$, and approximate the position and
velocity derivatives as
$\frac{dx(t)}{dt}\simeq\frac{x_{n+1}-x_{n}}{\epsilon}$ and
$\frac{dv(t)}{dt}\simeq\frac{v_{n+1}-v_{n}}{\epsilon}$, with
discrete-time position $x_n\equiv x(n\epsilon)$ and discrete-time
velocity $v_n\equiv v(n\epsilon)$. In this way, we turn the equations
of motion (\ref{eq:systemeom}) into the following Markov chain
\begin{equation}
\begin{cases}
    x_{n+1}=  x_n +  v_n \epsilon+\sqrt{2\epsilon D_F}~\xi_n\\
    v_{n+1}=-k\epsilon x_n+(1-\gamma \epsilon)v_n+\sqrt{2\epsilon D}~\eta_n 
    \end{cases},
    \label{eq:refdisc}
\end{equation}
where $\{\xi_n\}_{n\geq 0}$ and $\{\eta_n\}_{n\geq 0}$ are two
independent sequences of i.i.d. standard Gaussian random variables and
$D_F$ is a fictitious diffusion coefficient that goes to zero when
$\epsilon\downarrow 0$. The precise dependence of $D_F$ on $\epsilon$
is irrelevant, but we suppose that $D_F\propto\epsilon$ for
simplicity. Results for the original continuum system will be obtained
by performing the \textit{continuum limit} $\epsilon\downarrow 0$.
The drift and diffusion matrix extracted from Eq.\ (\ref{eq:refdisc})
are
\begin{equation}
S\equiv
\begin{pmatrix}
1 & \epsilon\\
-k \epsilon & 1-\gamma \epsilon
\end{pmatrix}\ 
\quad\text{and}\quad C\equiv
\begin{pmatrix}
\sqrt{2 \epsilon D_F } & 0\\
0 & \sqrt{2 \epsilon D}
\end{pmatrix}.
\label{eq:explSC}
\end{equation}
The eigenvalues of $S$ are
$1-\frac{\epsilon}{2}(\gamma\pm\sqrt{\gamma^2-4k})$, so that
$\rho(S)<1$ for all sufficiently small $\epsilon$. Thus, the
stationary covariance matrix exists for all sufficiently small
$\epsilon$ and reads
\begin{equation*}
\begin{aligned}
\Sigma_s&\equiv\sum_{m\geq0}S^m C^2(S^\top)^m\\
&=
\begin{pmatrix}
  \frac{2D(2-\gamma\epsilon+k\epsilon^2)+2D_F(2\gamma^2+2k-\gamma^3\epsilon-2\gamma k\epsilon+k\gamma^2\epsilon^2)}
       {k (\gamma-k \epsilon) \left(4-2 \gamma  \epsilon +k \epsilon ^2\right)}
  & -\frac{2D\epsilon+2D_F(2\gamma-\gamma ^2\epsilon-k\epsilon+\gamma k \epsilon^2)}{ (\gamma-k \epsilon) \left(4-2 \gamma  \epsilon +k \epsilon ^2\right)} \\
 -\frac{2D\epsilon+2D_F(2\gamma-\gamma ^2\epsilon-k\epsilon+\gamma k \epsilon^2)}{ (\gamma-k \epsilon) \left(4-2 \gamma  \epsilon +k \epsilon ^2\right)} &
 \frac{4D+ 2D_F(2k -\gamma k\epsilon+k^2\epsilon^2)}{(\gamma-k \epsilon) \left(4-2 \gamma  \epsilon +k \epsilon ^2\right)} \\
\end{pmatrix}\\
&=\begin{pmatrix}
\frac{D}{k\gamma}  & 0\\
0                   & \frac{D}{\gamma}
\end{pmatrix}+\mathcal O(\epsilon).
\end{aligned}
\end{equation*}
The leading order corresponds to the stationary covariance matrix for
the continuum system \cite{gardiner2009}, as expected.

Concerning the work injected by the random force into the Brownian
particle, the integral contribution in Eq.\ (\ref{eq:work1}) is
discretized according to the trapezoidal rule for convenience as
$\int_0^\tau v(t) \dot x(t) dt \simeq \frac{1}{2}\sum_{n=0}^{N-1}
(v_{n+1}+v_n)(x_{n+1}-x_n)$, so that $W_\tau\simeq W_N$ with the
discrete-time work
\begin{equation}
    W_N=\frac{1}{2}(v^2_N+kx^2_N+\gamma x_Nv_N)-\frac{1}{2}(v^2_0+kx^2_0+\gamma x_0 v_0)+\frac{\gamma}{2}\sum_{n=1}^N(v_{n-1}x_n- v_n x_{n-1}).
    \label{eq:work_discr}
\end{equation}
Once $W_N$ is recast as in Eq.\ (\ref{eq:quadfunct}), we find
\begin{equation}
-L=R\equiv
\begin{pmatrix}
    k  & \frac{\gamma}{2}  \\  
    \frac{\gamma}{2}  &  1
\end{pmatrix} , \quad \quad
U\equiv\begin{pmatrix}
    0  &  0  \\  
    0  &  0
\end{pmatrix} , \quad \quad
V\equiv
\begin{pmatrix}
    0  &  \frac{\gamma}{2}   \\  
    -\frac{\gamma}{2}   &  0
\end{pmatrix}. 
\label{eq:explRLVU}
\end{equation}
From Eq.\ (\ref{eq:mathave}), with full probability the typical value of
the discrete-time work turns out to be
\begin{equation*}
    \lim_{N\uparrow\infty}\frac{W_N}{N}=\frac{\gamma  \epsilon  (D+D_F k)}{\gamma -k \epsilon }=D\epsilon +\mathcal O(\epsilon^2) .
    \label{workave}
\end{equation*}
This formula and $\tau=N\epsilon$ give the typical value of the work
for the continuum system: $\braket{w}\equiv\lim_{\tau\uparrow\infty}W_\tau/\tau=D$.
To conclude, we observe that if
$J(w)\equiv-\lim_{N\uparrow\infty}\frac{1}{N}\log P(W_N/N=w)$ denotes
the rate function of $W_N/N$ in Eq.\ (\ref{eq:discrete_RF}), then the rate function $I(w)$ of
$W_\tau/\tau$ is
\begin{equation}
  I(w)= \lim_{\epsilon \downarrow 0}\frac{J(\epsilon w)}{\epsilon}=\lim_{\epsilon \downarrow 0}\sup_{\mu\in E}\bigg\{w\mu-\frac{\varphi(\mu)}{\epsilon}\bigg\}.
\label{eq:J2I}
\end{equation}

\subsection{Scaled cumulant generating function}
\label{sec:scgf}

Here we aim to evaluate the SCGF $\varphi(\mu)$ in Eq.\ (\ref{eq:J2I}).
The first step is the knowledge of the Hermitian matrix
$F_\mu(\theta)$ defined by Eq.\ (\ref{eq:mathsymbol}). From
Eq.\ (\ref{eq:explSC}) and Eq.\ (\ref{eq:explRLVU}), we find
\begin{equation*}
    F_\mu(\theta)=
    \begin{pmatrix}
    l_{11}+m_{11}\cos \theta & l_{12}+m_{12}\cos \theta-\imath p_{12}\sin \theta\\
    l_{12}+m_{12}\cos \theta+\imath p_{12}\sin \theta & l_{22}+m_{22}\cos \theta
    \end{pmatrix},
\end{equation*}
where the explicit expressions of the coefficients are
\begin{equation*}
\begin{aligned}
  l_{11}&\equiv\frac{1}{ D_F \epsilon }+\frac{k^2  }{2 D}\epsilon  , \quad m_{11}\equiv-\frac{1}{ D_F \epsilon } ,\\
l_{12}&\equiv-\frac{k}{2 D}+\frac{1}{2  D_F}+\frac{\gamma  k  }{2 D}\epsilon , \quad   m_{12}\equiv\frac{k}{2 D}-\frac{1}{2  D_F}  ,
   \quad  p_{12}&\equiv\gamma  \mu +\frac{k}{2D}+\frac{1}{ 2D_F} ,\\
 l_{22}&\equiv\frac{1}{D \epsilon }-\frac{\gamma }{D}+  \left(\frac{\gamma ^2}{2 D}+\frac{1}{2  D_F}\right)\epsilon  ,
  \quad m_{22}\equiv\frac{\gamma }{D} .
\end{aligned}
\end{equation*}
We note that only the coefficient $p_{12}$ depends on the variable
$\mu$.  The determinant of $F_\mu(\theta)$ is given by
\begin{equation*}
    \det F_\mu(\theta)=\frac{1}{DD_F}\left(a_0+a_1\cos\theta+a_2\cos^2\theta\right),
\end{equation*}
where
\begin{equation*}
  \begin{aligned}
     a_0&\equiv\frac{1}{\epsilon ^2}-\frac{\gamma }{\epsilon }-D D_F\gamma ^2 \mu ^2-D_F\gamma  k \mu -D\gamma  \mu
     +\frac{\gamma ^2}{2}-\frac{\gamma  k  }{2 }\epsilon+\frac{k^2 }{4}\epsilon ^2,\\
     a_1&\equiv-\frac{2}{ \epsilon ^2}+\frac{2 \gamma }{ \epsilon }-k-\frac{\gamma ^2}{2}+\frac{\gamma  k }{2}\epsilon,\\
    a_2&\equiv\frac{1}{\epsilon ^2}-\frac{\gamma }{\epsilon }+D  D_F \gamma ^2 \mu ^2+k+\gamma  k \mu D_F+\gamma  \mu D  .
\end{aligned}
\end{equation*}

According to Sylvester's criterion, the Hermitian matrix
$F_\mu(\theta)$ is positive definite if and only if its upper left
entry and its determinant are positive. The former is positive for all
$\theta$, $\mu$, and $\epsilon>0$. For $\epsilon>0$ small enough, the
latter is positive for all $\theta$ provided that
$\mu\in(\tilde\mu_-,\tilde\mu_+)$ with
\begin{equation*}
    \tilde\mu_\pm\equiv\frac{-(D+D_Fk)\pm\sqrt{D D_F\gamma^2+(D+D_Fk)^2+D_F D k(k\epsilon-2\gamma)\epsilon}}{2 D D_F \gamma} .
\end{equation*}
Therefore, recalling the discussion in Subsection
\ref{sec:anal_overview}, the primary domain of the SCGF $\varphi(\mu)$
is $O=(\tilde\mu_-,\tilde\mu_+)$.  Bearing in mind that
$D_F\propto\epsilon$, we note that
\begin{equation}
    \begin{split}
        \tilde\mu_+&=\frac{\gamma}{4D}+\mathcal{O}(\epsilon),\\
        \tilde\mu_-&=-\frac{1}{D_F\gamma}+\mathcal{O}(\epsilon^0).
    \end{split}
    \label{eq:mu_primary_expansion}
\end{equation}
We can now compute the function $\varphi(\mu)$. For $\mu\in O$, we
have the identity
\begin{equation}
\frac{1}{2\pi}\int_0^{2\pi}d\theta~\log\big(a_0+a_1\cos\theta+a_2\cos^2\theta\big)=2\log\left(\frac{g_1+g_2+\sqrt{2(a_0-a_2+g_1g_2)}}{4}\right) 
 \label{eq:int1_marco}
\end{equation}
with 
\begin{equation}
    \begin{split}
        g_1&\equiv\sqrt{a_0-a_1+a_2}=\frac{2}{\epsilon}-\gamma +\frac{k \epsilon}{2},\\
        g_2&\equiv\sqrt{a_0+a_1+a_2}=\frac{k \epsilon}{2}.
    \end{split}
    \label{eq:scgf_solution}
\end{equation}
Thus, combining Eq.\ (\ref{eq:math_SCGF}) with Eq.\ (\ref{eq:int1_marco}),
we get the explicit expression
\begin{align}
  \nonumber
  \varphi(\mu)&=-\log\left(\frac{g_1+g_2+\sqrt{2(a_0-a_2+g_1g_2)}}{4}\right) -\log(2\epsilon)\\
  \nonumber
  &=\frac{1}{2} \Big[\gamma -\sqrt{\gamma(\gamma-4D\mu)}\Big]\epsilon+\mathcal O(\epsilon^2).
\end{align}
We define
\begin{equation}
    \phi(\mu)\equiv\lim_{\epsilon \downarrow 0}\frac{\varphi(\mu)}{\epsilon}=
\frac{\gamma}{2}\left(1-\sqrt{1-\frac{4D\mu}{\gamma}}\right),
\label{eq:scgf_final}
\end{equation}
to which we refer as the SCGF for the continuum system. This function
is defined on the primary domain $(-\infty,\gamma/4D)$, which,
manifestly, is the continuum limit of
Eq.\ (\ref{eq:mu_primary_expansion}).

\subsection{Effective domain}
\label{sec:eff_dom}

Here we establish the effective domain $E$ in Eq.\ (\ref{eq:J2I}). To
this aim, the explicit expressions of the matrices $\mathcal L_\mu$
and $\mathcal R_\mu$ in Eq.\ (\ref{eq:mathLR}) are needed. This requires
at first to compute the matrices $\Phi_\mu(0)$ and $\Phi_\mu(1)$ in
Eq.\ (\ref{eq:phiexp}) for $\mu\in O$. All the involved integrals in
$\Phi_\mu(0)$ and $\Phi_\mu(1)$ fall within the following general case:
\begin{equation*}
\begin{aligned}
\frac{1}{2\pi}&\int_0^{2\pi} d\theta~\frac{\zeta_0+\zeta_1\cos\theta+\zeta_2\cos^2\theta}{a_0+a_1\cos\theta+a_2\cos^2\theta}=\\
=&\frac{(a_2\zeta_0-\zeta_2a_0)(g_1+g_2)+(a_2\zeta_1-\zeta_2a_1)(g_1-g_2)}{a_2g_1g_2 \sqrt{2(a_0-a_2+g_1g_2})}
+ \frac{\zeta_2}{a_2},
\end{aligned}
\end{equation*}
where $g_1$ and $g_2$ are as in Eq.\ (\ref{eq:scgf_solution}), whereas
$\zeta_0$, $\zeta_1$, and $\zeta_2$ are generic numbers. Thus, for
$\mu\in O$, we find
\begin{align}
  \nonumber
  \Phi_\mu(0)&=\frac{DD_F(g_1+g_2)}{g_1g_2 \sqrt{2(a_0-a_2+g_1g_2})}
\begin{pmatrix}
 l_{22} & -l_{12} \\
 -l_{12} & l_{11} \\
\end{pmatrix}\\
\nonumber
&+\frac{DD_F(g_1-g_2)}{g_1g_2 \sqrt{2(a_0-a_2+g_1g_2})}
\begin{pmatrix}
 m_{22} & -m_{12} \\
 -m_{12} & m_{11} \\
\end{pmatrix},\\
  \nonumber
  \Phi_\mu(1)&=\frac{DD_F(g_1-g_2)}{g_1g_2 \sqrt{2(a_0-a_2+g_1g_2})}
\begin{pmatrix}
 l_{22} & -l_{12} \\
-l_{12} & l_{11} \\
\end{pmatrix}\\
\nonumber
&+\frac{DD_F}{a_2}\bigg\{1-\frac{a_0(g_1+g_2)+a_1(g_1-g_2)}{g_1g_2 \sqrt{2(a_0-a_2+g_1g_2})}\bigg\}
\begin{pmatrix}
 m_{22} & -m_{12} \\
-m_{12} & m_{11} \\
\end{pmatrix}\\
\nonumber
&+\frac{DD_F}{a_2}\bigg\{1-\frac{(a_0+a_2)(g_1+g_2)+a_1(g_1-g_2)}{g_1g_2 \sqrt{2(a_0-a_2+g_1g_2})}\bigg\}
\begin{pmatrix}
0 & - p_{12} \\
p_{12} & 0 \\
\end{pmatrix}.
\end{align}
Combining these expressions with Eq.\ (\ref{eq:mathHK}) and
Eq.\ (\ref{eq:mathLR}), expanding around $\epsilon=0$ with
$D_F\propto\epsilon$, a lengthy calculation that we omit gives the
following simple result for the leading order of the matrices
$\mathcal L_\mu$ and $\mathcal R_\mu$:
\begin{equation*}
    \begin{aligned}
        \mathcal L_\mu&=
       \begin{pmatrix}
 k\frac{-\gamma +\sqrt{\gamma  (\gamma -4 D \mu )}+2 D \mu }{2 D}+\frac{1}{\sigma^2_x} & 0 \\
 0 & \frac{-\gamma +\sqrt{\gamma  (\gamma -4 D \mu )}+2 D \mu }{2 D}+\frac{1}{\sigma^2_v} \\
 \end{pmatrix} 
+\mathcal O(\epsilon),\\
\mathcal R_\mu&=
    \frac{\gamma +\sqrt{\gamma  (\gamma -4 D \mu )}-2 D \mu }{2 D}
\begin{pmatrix}
 k & 0  \\
 0 & 1 \\
\end{pmatrix} 
+\mathcal O(\epsilon).
    \end{aligned}
\end{equation*}

The interval $E\equiv(\mu_-,\mu_+)$ accounts for those $\mu\in O$ that
make $\mathcal{L}_\mu$ and $\mathcal{R}_\mu$ positive-definite at the
same time.  It is straightforward to check that, at small $\epsilon$,
the matrix $\mathcal R_\mu$ is positive definite for all $\mu$ in the
primary domain $O$.  Instead, at small $\epsilon$, the matrix
$\mathcal L_\mu$ is positive definite only if both the diagonal
entries of the leading order are positive, that is only if
\begin{equation}
 \sqrt{\gamma  (\gamma -4 D \mu )}+2 D \mu >\gamma-\frac{2D}{\max\{k\sigma^2_x,\sigma^2_v\}}.
\label{eq:effective_constraint}
\end{equation}
In this way, the effective domain of the SCGF in the continuum limit,
which is $\phi(\mu)$ given by Eq.\ (\ref{eq:scgf_final}), is the set
of numbers $\mu<\gamma/4D$ satisfying
Eq.\ (\ref{eq:effective_constraint}).  Solving
Eq.\ (\ref{eq:effective_constraint}), for the continuum system we
finally find $E=(\mu_-,\mu_+)$ with
\begin{equation}
\begin{aligned}
  \mu_-&=-\frac{1}{M}-\sqrt{\frac{\gamma}{DM}},\quad M\equiv\max\big\{k\sigma_x^2,\sigma_v^2\big\},\\
  \mu_+&=\begin{cases}
   \frac{\gamma}{4D} &\mbox{if } M\le 4D/\gamma\\
   -\frac{1}{M}+\sqrt{\frac{\gamma}{DM}} &\mbox{if } M> 4D/\gamma\\
\end{cases}.
\end{aligned}
\label{eq:mu_effective}
\end{equation}
We note that $\mu_->\tilde\mu_-=-\infty$ for every choice of the
parameters, so that $\phi(\mu)$ is non-steep at $\mu_-$ in any
situation, except for the limiting case of concentrated initial values
defined by the limits $\sigma_x,\sigma_v\downarrow 0$. With
$\phi(\mu)$ given by Eq.\ (\ref{eq:scgf_final}) and $\mu_\pm$ given by
Eq.\ (\ref{eq:mu_effective}), from Eq.\ (\ref{eq:J2I}) we get
\begin{equation}
  I(w)=\sup_{\mu\in(\mu_-,\mu_+)}\big\{w\mu-\phi(\mu)\big\}.
\label{eq:I_final}
\end{equation}

\subsection{Rate function}
\label{sec:sec_rf}

Here we investigate $I(w)$ given by Eq.\ (\ref{eq:I_final}), which
constitutes the rate function for the continuum system and is one of the main
objectives of the paper. Using Eq.\ (\ref{eq:scgf_final}) for
$\phi(\mu)$ and Eq.\ (\ref{eq:mu_effective}) for $\mu_\pm$, when
$M\equiv\max\{k\sigma_x^2,\sigma_v^2\}\le 4D/\gamma$ we find
\begin{equation*}
  I(w)=\begin{cases}
  w\mu_-+\phi(\mu_-)= \sqrt{\frac{D\gamma}{M}}-w\Big(\frac{1}{M}+\sqrt{\frac{\gamma}{DM}}\Big) & \text{if~} w\le w_-\\
   \frac{\gamma}{4Dw}(w-D)^2 &\text{if~} w>w_-
  \end{cases}
\end{equation*}
with
\begin{equation*}
    w_-\equiv\lim_{\mu\downarrow\mu_{-}}\phi'(\mu)=\frac{D}{\sqrt{1-\frac{4D}{\gamma}\mu_-}}=\frac{D}{1+\sqrt{\frac{4D}{\gamma M}}}>0.
\end{equation*}
If instead $M\equiv\max\{k\sigma_x^2,\sigma_v^2\}>4D/\gamma$, then we
get
\begin{equation*}
  I(w)=\begin{cases}
  w\mu_--\phi(\mu_-)=\sqrt{\frac{D\gamma}{M}}-w\Big(\frac{1}{M}+\sqrt{\frac{\gamma}{DM}}\Big) & \text{if~} w\le w_-\\
   \frac{\gamma}{4Dw}(w-D)^2 &\text{if~} w_-<w<w_+\\
  w\mu_+-\phi(\mu_+)= w\Big(\sqrt{\frac{\gamma}{DM}}-\frac{1}{M}\Big)-\sqrt{\frac{D\gamma}{M}}  & \text{if~} w\ge w_+
  \end{cases} ~,
\end{equation*}
where $w_-$ is as before and
\begin{equation*}
    w_+\equiv\lim_{\mu\uparrow \mu_{+}}\phi'(\mu)=\frac{D}{\sqrt{1-\frac{4D}{\gamma}\mu_+}}=\frac{D}{1-\sqrt{\frac{4D}{\gamma M}}}>w_-.
\end{equation*}
We realise that the rate function always has a singularity at $w_-$ below the
mean $D$ with an associated left linear tail. Moreover, when
$M>4D/\gamma$ the model exhibits a second singularity at $w_+$ above
$D$ and a corresponding right linear tail.  We stress that the
parameter $M$ is the only place where the elastic constant $k$
appears.

\begin{figure}[t]
    \centering
    {\includegraphics[width=1\textwidth]{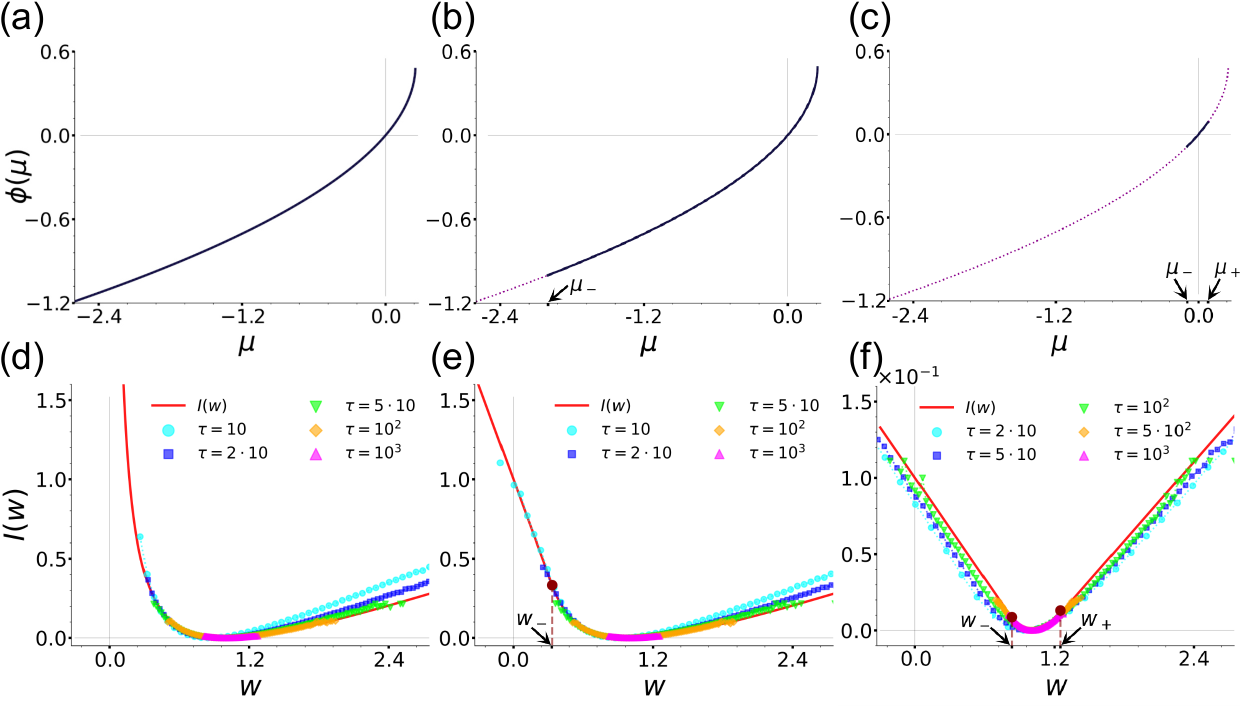}}
    \caption{\footnotesize{SCGF $\phi(\mu)$ (top row) and rate function $I(w)$
        (bottom row) with system parameters $\gamma=1.0$, $D=1.0$, and
        $k=1.0$ (in arbitrary units) for several initial
        conditions. $(a)$ and $(d)$: concentrated initial values
        $x(0)=v(0)=0$ given by the limits $\sigma_x,\sigma_v\downarrow
        0$. $(b)$ and $(e)$: stationary system corresponding to
        $\sigma_x^2=D/k\gamma$ and $\sigma_v^2=D/\gamma$.  $(c)$ and
        $(f)$: generic initial conditions defined by $\sigma_x=10$ and
        $\sigma_v=10$ (in arbitrary units). In the top panels, the
        blue line shows the SCGF in the effective domain, whereas the
        purple dashed curve in $(b)$ and $(c)$ depicts its trend in
        the primary domain (see Subsections \ref{sec:scgf} and
        \ref{sec:eff_dom} for details). In the bottom panels, the red
        solid line is the analytical rate function, whereas the coloured dotted
        lines are the numerical rate functions at different values of $\tau$. In
        $(e)$ and $(f)$, the red dashed lines mark the beginning of the 
        left linear tail at $w_-$ and of the right linear tail at $w_+$.
    }}
\label{fig:rf_comparison}
\end{figure}

In the special case of concentrated initial values defined by the
limits $\sigma_x,\sigma_v\downarrow 0$, i.e., when the Brownian
particle starts fluctuating from $x(0)=v(0)=0$, we have $M=0\le
4D/\gamma$, so that the rate function $I(w)$ reads
\begin{equation}
    I(w)=\begin{cases}
        +\infty         &\text{if~}w\leq 0\\
        \frac{\gamma}{4Dw}(w-D)^2      &\text{if~}w>0
    \end{cases} .
    \label{eq:rf_no_staz}
\end{equation}
In this limiting case the rate function does not display linear tails since the
assumption $x(0)=v(0)=0$ rules out the possibility of negative
fluctuations for the work, as it becomes a sum of positive
contributions according to Eq.\ (\ref{eq:work1}). We note that the rate function
is independent of the elastic constant $k$ in this limiting case.  The
rate function is independent of $k$ also in the case of stationary system,
corresponding to $\sigma^2_x=D/k\gamma$ and $\sigma^2_v=D/\gamma$. In
fact, for a stationary system we find $M=D/\gamma\le 4D/\gamma$,
giving
\begin{equation}
    I(w)=\begin{cases}
        \frac{\gamma}{D}(D-2w)         &\text{if~}w\leq \frac{D}{3}\\
        \frac{\gamma}{4Dw}(w-D)^2      &\text{if~}w> \frac{D}{3}
    \end{cases} .
    \label{eq:rf_staz}
\end{equation}
The left linear tail, and that alone, is now present.
Eqs.\ (\ref{eq:rf_no_staz}) and (\ref{eq:rf_staz}) demonstrate the
claim made in Ref.\ \cite{farago2002} that the injected power is
completely insensitive to the presence of a harmonic confinement under
concentrated initial values or stationary initial conditions.
Consistently, they reproduce the analytic results of
Ref.\ \cite{farago2002} for the free system corresponding to $k=0$.
As observed in Ref.\ \cite{farago2002}, the rate function satisfies no
Fluctuation Relation \cite{Gallavotti_1995, Kurchan_1998,
  Lebowitz_1999, seifert2012}.
  
  \begin{figure}[t]
    \centering
    {\includegraphics[width=0.95\textwidth]{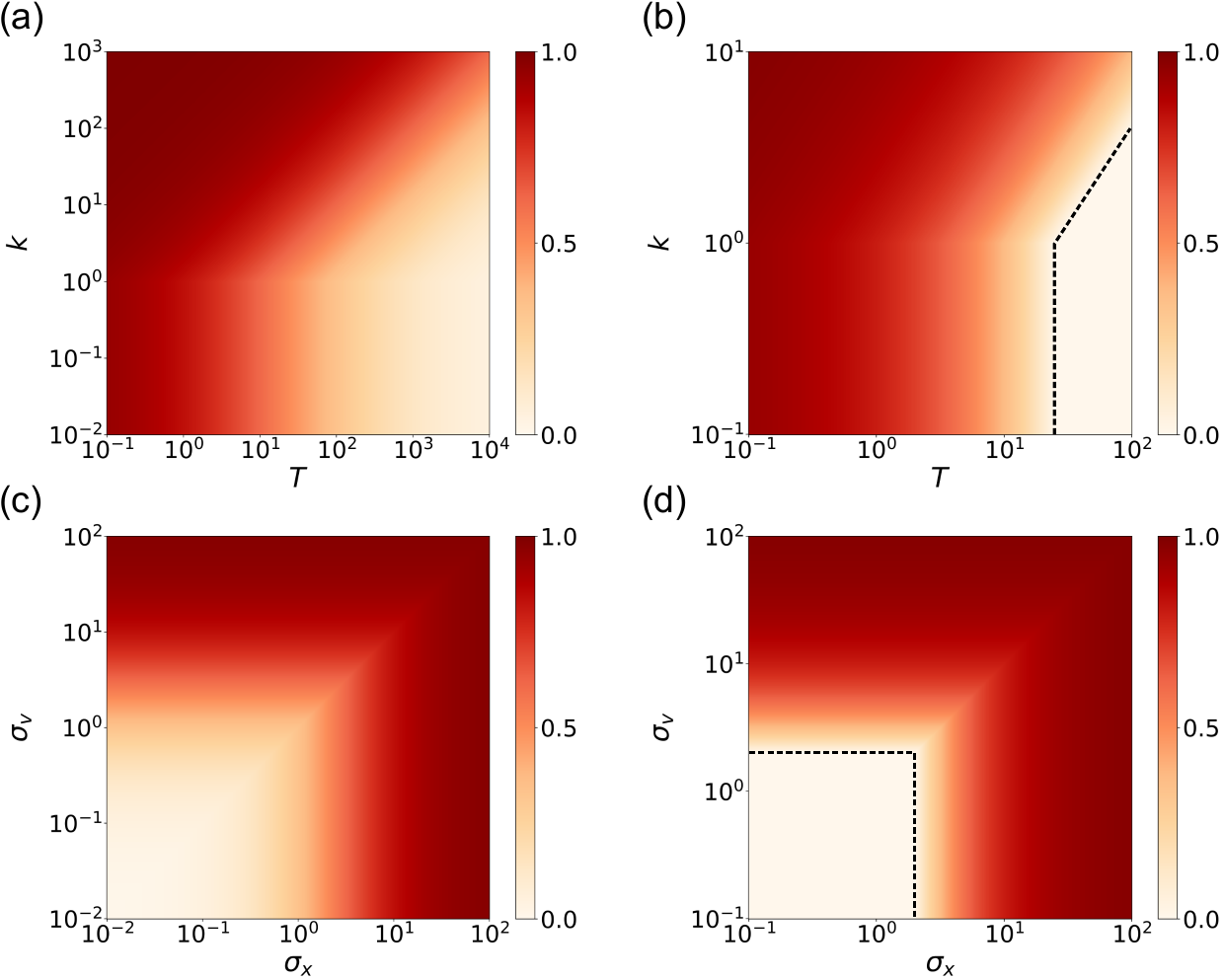}}
    \caption{\footnotesize{Phase diagrams as deduced by the ratios
        $r_-\equiv w_-/D<1$ and $r_+\equiv D/w_+<1$ between the rate function
        singularities $w_\pm$ and the mean value of the work
        $D$. Regarding $r_+$, white regions delimited by black dashed
        lines are the regions of parameters where the right linear
        tail does not occur, that is where
        $M\equiv\max\{k\sigma_x^2,\sigma_v^2\}\le 4D/\gamma$.  $(a)$
        and $(b)$: $r_-$ and $r_+$ for $D=\gamma k_BT$ in the $T-k$
        plane with $\gamma=1.0$, $k_B=1.0$, $\sigma_x=10$ and
        $\sigma_v=10$ (in arbitrary units). $(c)$ and $(d)$: $r_-$ and
        $r_+$ in the $\sigma_x-\sigma_v$ plane with $\gamma=1.0$,
        $D=1.0$ and $k=1.0$ (in arbitrary units).}}
\label{fig:eff_bounds}
\end{figure}

Fig.\ \ref{fig:rf_comparison} reports the SCGF $\phi(\mu)$ (top) and
the rate function $I(w)$ (bottom) for the concentrated initial values
$x(0)=v(0)=0$, for stationary initial conditions and for generic
initial conditions entailing both left and right linear tails. The
figure summarises the ability of the model to exhibit no linear tail,
one linear tail, or two linear tails. In order to confirm analytical
results, Fig.\ \ref{fig:rf_comparison}(bottom) also compares the
analytical rate function with rate functions estimated by means of numerical simulations of
the original model. In fact, we simulated the trajectories of the
discrete-time model (\ref{eq:refdisc}) with time step
$\epsilon=10^{-2}$ and fictitious diffusion coefficient $D_F$ set
equal to 0. Then, we sampled the injected work $W_\tau$ for different
times $\tau=N\epsilon$ as prescribed by Eq.\ (\ref{eq:work_discr}).
In the regions of $w$ around $D$ that we are able to explore
numerically, we find a good agreement between analytic results and
simulations. Recall that $D$ is the typical value of $W_\tau$, with
$D=1.0$ (in arbitrary units) in Fig.\ \ref{fig:rf_comparison}.

Fig.\ \ref{fig:eff_bounds} provides some phase diagrams of the system
as deduced by the ratios $r_-\equiv w_-/D$ and $r_+\equiv D/w_+$
between the rate function singularities $w_\pm$, where linear tails begin, and
the mean work $D$. We have set $r_+\equiv0$ when the right linear tail
does not occur, that is when $M\equiv\max\{k\sigma_x^2,\sigma_v^2\}\le
4D/\gamma$. In this way, we have $r_-\in(0,1)$ and $r_+\in[0,1)$, with
  $r_-\uparrow 1$ when $w_-$ approaches $D$ from below, $r_+\uparrow
  1$ when $w_+$ approaches $D$ from above, and $r_+\downarrow 0$ when
  the region $M\le 4D/\gamma$ is approached from outside. Panels (a)
  and (b) of Fig.\ \ref{fig:eff_bounds} report the phase diagrams for
  $r_-$ and $r_+$, respectively, in the $T-k$ plane at fixed $\gamma$,
  $\sigma_x$ and $\sigma_v$ with $D=\gamma k_BT$.  We see that $w_-$
  and $w_+$ approach $D$ at low temperature $T$ and/or large elastic
  constant $k$. At higher $T$ and finite $k$, $w_-$ decreases to zero,
  whereas $w_+$ increases to infinity up to a critical value of $T$
  where the right linear tail ceases to exist. This critical value is
  identified by the condition $4k_BT=\max\{k\sigma_x^2,\sigma_v^2\}$.
  Panels (c) and (d) of Fig.\ \ref{fig:eff_bounds} report the phase
  diagrams for $r_-$ and $r_+$ in the $\sigma_x-\sigma_v$ plane at
  fixed system parameters. We note that both $w_-$ and $w_+$ approach
  $D$ at large $\sigma_x$ and/or large $\sigma_v$. On the contrary,
  when $\sigma_x$ and $\sigma_v$ become small, the point $w_-$
  decreases to zero, whereas $w_+$ ceases to exist as soon as
  $\max\{k\sigma_x^2,\sigma_v^2\}= 4D/\gamma$.

\section{Phenomenology of the trajectories}
\label{sec:sec-sings}

\begin{figure}[t]
    \centering
    {\includegraphics[width=1\textwidth]{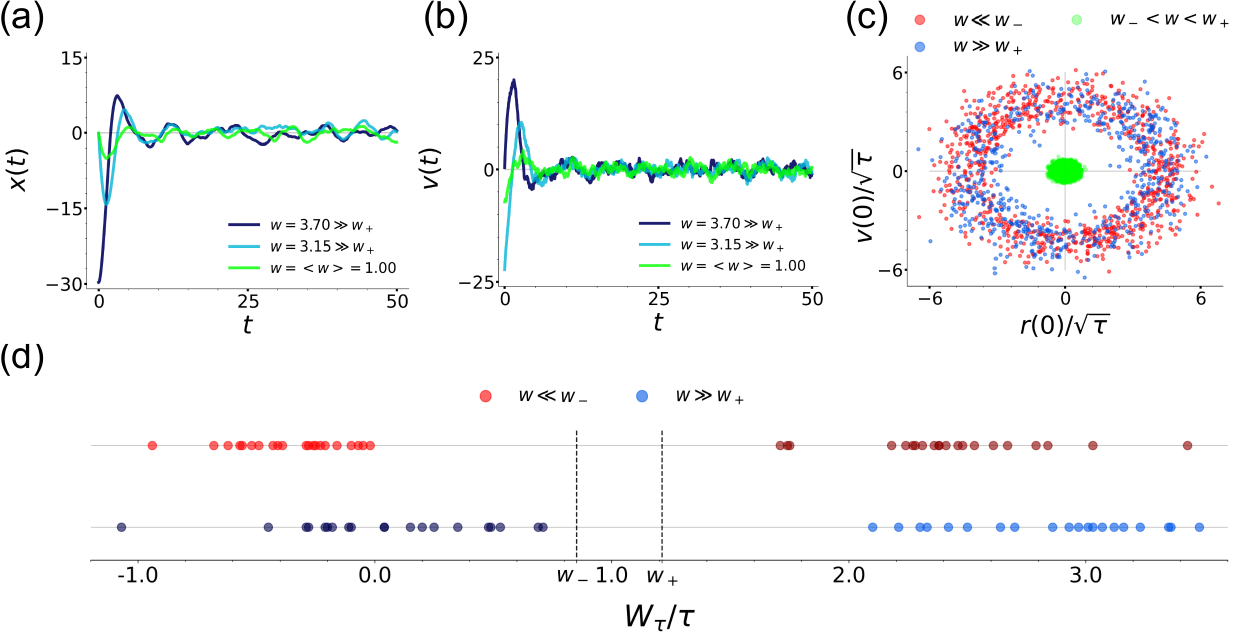}}
    \caption{\footnotesize{Trajectory characterisations. $(a)$ and
        $(b)$: typical position and velocity trajectories under
        non-stationary initial conditions corresponding to the works
        $w=3.70,~3.15\gg w_+=1.25$ (blue) and $w\sim \braket{w}=D=1.0$
        (green). $(c)$: initial position and velocity for a pool of
        $10^7$ trajectories conditional on $W_\tau=w\tau$ with
        $w\leq-0.6 \ll w_-=0.83$ (red), $w\geq3.2\gg w_+=1.25$ (blue)
        and $w_-\leq w\leq w_+$ (green). $(d)$: comparison between the
        values $w$ of the work per unit time generated by singular
        trajectories ($w\ll w_-$ red, $w\gg w_+$ blue) and the values
        of the work per unit time associated with same initial
        conditions but opposite random force path (corresponding
        darker colours). The dashed vertical lines highlight the
        locations of $w_\pm$.  In all cases we fixed $\gamma=1.0$,
        $D=1.0$, $k=1.0$, $\sigma_x=10$, $\sigma_v=10$ and
        $\tau=50\gg t_I=1.0$ (in arbitrary units).}}
\label{fig:phase_space}
\end{figure}

%\textcolor{red}{In Ref.\ \cite{farago2002}, the physical origin of the
 % singularity in the rate function of the stationary, free particle
 % was ascribed to a very energetic initial condition.  In this section
 % we corroborate this hypothesis by exploring the mechanisms that
 % originate singular trajectories, i.e., trajectories giving values
 % $w$ of the work in a linear tail of the rate function $I(w)$.} 
    
In this section we shed light on the physical mechanisms that
originate singular trajectories, i.e.\ trajectories giving values $w$
of the work in a linear tail of the RF $I(w)$.
We pursue the goal by analysing the trajectories of the model with
system parameters as in Fig.\ \ref{fig:rf_comparison}f, so that the
rate function has both left and right linear tails. The simulation
time $\tau$ we consider is much larger than the inertial time
$t_I\equiv\gamma^{-1}$. We fixed $\gamma=1.0$, $D=1.0$, $k=1.0$,
$\sigma_x=10$, $\sigma_v=10$, and $\tau=50\gg t_I=1.0$ (in arbitrary
units).

Inspection of singular trajectories reveals the presence of some big
jump phenomena. Fig.\ \ref{fig:phase_space} reports three trajectories
of the system, in terms of the position $x(t)$ in
Fig.\ \ref{fig:phase_space}a and the velocity $v(t)$ in
Fig.\ \ref{fig:phase_space}b, conditional on $W_\tau=w\tau$ with one
value $w\sim \braket{w}=D$ (green lines) and two values $w\gg w_+$
(blue lines) in the far right linear tail.  We see that the singular
trajectories are characterised by a short initial relaxation transient
due to a big jump in one of the initial values: $x(0)\sim -30$ in
Fig.\ \ref{fig:phase_space}a for $w=3.70$ and $v(0)\sim-25$
Fig.\ \ref{fig:phase_space}b for $w=3.15$. An analogous behaviour
characterises also the trajectories corresponding to $w\ll w_-$ in the
far left linear tail. Such big jumps are confirmed and characterised
by Fig.\ \ref{fig:phase_space}c, where the initial position and
velocity are reported for a large pool of trajectories conditional on
$W_\tau=w\tau$ with $w\ll w_-$ (red), $w\gg w_+$ (blue) and $w_-\leq
w\leq w_+$ (green). This panel clearly shows two different patterns
for singular and non-singular trajectories, the former roughly
localising on an annulus, so that at least one among $x(0)$ and $v(0)$
makes a big jump, the latter concentrating around the origin. At
larger times $\tau$, green points tend to accumulate more and more
densely around the origin, whereas red and blue points remain
essentially unaltered. This implies that big jumps in the initial
conditions are of order $\sqrt{\tau}$. Therefore, we conclude that big
jumps of order $\sqrt{\tau}$ in the initial values of position and
velocity are an essential ingredient for singular trajectories to
occur. Coherently with what we find here, we note
  that in Ref.\ \cite{farago2002} the physical origin of the
  singularity in the rate function of the stationary, free particle
  was ascribed to a very energetic initial condition.  Similarly, a
  big jump mechanism causing a particle to roll uphill in the
  potential was suggested in Ref.\ \cite{semeraro2023} to explain a
  singular rate function.

 % Moreover we remark
 % that, similarly to the big jumps mechanism, it has been shown that
 % particles rolling uphill in the potential cause a huge energy
 % dissipation in the system \cite{dabelow2021}, usually leading to
 % singularities in the rate function, resulting in turn in the
 % breakdown of the fluctuation theorem.

It is interesting to observe that the patterns for $w\ll w_-$ and
those for $w\gg w_+$ in Fig.\ \ref{fig:phase_space}c overlap, meaning
that same initial big jumps can produce trajectories associated to
values of the work in both the left and the right linear tail of the
rate function. In order to distinguish such trajectories in terms of the injected
work, we discuss the action of the random force $\eta(t)$.  In fact,
once a singular initial configuration is given, the linear tail where
the value of the work falls is determined by the sign of the random
force realisation, as demonstrated by
Fig.\ \ref{fig:phase_space}d. Specifically, consider a trajectory with
$w\gg w_+$. This trajectory involves big jumps in the initial
conditions and is characterised by a certain random force path.
Fig.\ \ref{fig:phase_space}d shows that if we now construct a
trajectory with same initial conditions and opposite random force,
then the corresponding value of the work falls in the left linear
tail. Similarly, if we consider a trajectory with $w\ll w_-$ and
change the sign of the random force, then we end up with a value of
the work in the right linear tail of the rate function. In conclusion, changing
the sign of the random force systematically turns a trajectory with
$w\gg w_+$ into a trajectory with $w\ll w_-$, and vice versa.

\section{Conclusions}
\label{sec:conclusions}

In this paper we have studied the large deviations of the power 
injected  by the random force of the thermal bath on a Brownian particle
under the action of an external harmonic potential with generic
uncorrelated Gaussian initial conditions.

In particular, we have computed analytically the SCGF and the rate
function using the approach recently proposed in
Ref.\ \cite{zamparo2023} for quadratic functionals of stable
Gauss-Markov chains. The SCGF is found to be non-steep at the left
boundary, except from the limiting case of concentrated initial
values, i.e., $x(0)=v(0)=0$. It is also non-steep at the right
boundary when the variance of the initial position or initial velocity
is large enough.  The corresponding rate function therefore can
exhibit no singularity and no linear tail, or one second-order
singularity and one linear tail on the left, or two second-order
singularities and two linear tails. Interestingly, the dependence of
the rate function on the elastic constant $k$ is found
  for large enough initial variances and lost for concentrated
initial values and stationary initial condition, thus confirming the
claim made in Ref.\ \cite{farago2002} for a confined Brownian
particle.

In order to understand the physical mechanism originating the
singularities, we have analysed the singular trajectories, i.e., the
trajectories that give values of the work either in the left or in the
right tail of the rate function. By resorting to numerical simulations, we have
shown that singular trajectories are characterised by big jumps of
order $\sim\sqrt{\tau}$ in the initial position and/or the initial
velocity. Once the big jumps occur, the random force path determines
if the value of the work falls within the left linear tail or the
right linear tail.

A natural extension of this research will be to
  consider correlations in the initial condition or coloured thermal
  noise, such as the Ornstein-Uhlenbeck process \cite{arnab2014, semeraro2021} or
  the fractional Brownian motion \cite{guggenberger2022}. Leaving the
  field of Gaussian processes, another possible extension will be to
  consider an anharmonic potential, which could disrupt the
  singularities of the rate function, and continuous-time random
  walks, for which large deviation principles are known in great
  generality \cite{ldp_rr_1,ldp_rr_2}.  On a more practical note, in
  analogy with experiments on Fluctuation Relations
  \cite{wang2002,gomez2010}, we argue that our findings could be
  tested by means of optical trap experiments with $\mu$m diameter
  beads.

\section{Acknowledgments}
This work has been supported by the Italian Ministry of University and
Research via the project PRIN/2020 PFCXPE and by Apulia Region via the
project UNIBA044 of the research programme REFIN - Research for
Innovation. GBC is financially supported by Apulia Region via the
initiative \textit{Dottorati di ricerca in Puglia XXXVII ciclo}.

\section*{References}
\bibliographystyle{IEEEtran} 
\bibliography{Bibliography.bib}

%\bibliographystyle{iopart-num.bst}		% The reference style
%\bibliography{Bibliography.bib} 	        % Multiple bib files

\end{document}